# Unexpected Structures for Intercalation of Sodium in Epitaxial Graphene-SiC Interfaces

Andreas Sandin[1], Thushari Jayasekera[2], J. E. (Jack) Rowe[1], Ki Wook Kim[3]
M. Buongiorno Nardelli[1,4] Daniel B. Dougherty[1]*
1. Department of Physics, North Carolina State University, Raleigh, NC 27695-8202
2. Department of Physics, Southern Illinois University-Carbondale, Carbondale, IL 62901-4401
3. Department of Electrical and Computer Engineering, North Carolina State University, Raleigh, NC 27695-7911
4. Computer Science and Mathematics Division, Oak Ridge National Laboratory, Oak Ridge, TN 37831-6359

We show using scanning tunneling microscopy, spectroscopy, and *ab initio* calculations that several intercalation structures exist for Na in epitaxial graphene on SiC(0001). Intercalation takes place at room temperature and Na electron-dopes the graphene. It intercalates in-between single-layer graphene and the carbon-rich interfacial layer. It also penetrates beneath the interfacial layer and decouples it to form a second graphene layer. This decoupling is accelerated by annealing and is verified by direct Na deposition onto the interface layer. Our observations show that intercalation in graphene is *fundamentally different than in graphite* and is a versatile means of electronic control.

Graphene is the most promising electronic material to be discovered in the past decade.[1] Its high carrier mobility and chemical and mechanical robustness suggest important applications in electronics that have already started to be realized in prototype devices.[2] A crucial goal in ongoing graphene research is to find ways to control its physical properties by chemical doping.

One strategy for controlling the physical properties of graphene has been adapted from the field of graphite intercalation compounds.[3] It is possible to insert impurity atoms in between sheets of graphite, where they become chemically bound. The most striking example is calcium intercalation in graphite which results in a superconducting solid ($CaC_6$) with a transition temperature of above 11 K.[4] Unfortunately, intercalation processes in graphite are often inhibited by slow kinetics.[3,5] Intercalation beneath *graphene* may be more efficient since intercalating species have only to be inserted beneath a single layer or perhaps a few atomic layers.

Several atomic species have been intercalated beneath epitaxial graphene grown on both SiC(0001) and metallic substrates. On the Si terminated SiC(0001) surface, graphene is allowed to grow nearly decoupled from substrate supported by an interfacial carbon rich layer (aka "buffer layer") that can be seen as a covalently bonded graphene-like sheet[6] with a large bandgap and vanished Dirac cone band structure[7]. Recent studies[8-10] demonstrated that this buffer layer could be "activated" by intercalation of hydrogen atoms at high temperatures. Hydrogen intercalation breaks covalent bonds between the buffer layer and the SiC substrate to transform this interfacial layer into a purely $sp^2$ bonded sheet of graphene.

Other experimental studies have demonstrated intercalation structures for oxygen[11] and for compounds that can alternate the carrier concentration in the graphene. Fluorine intercalation gives p-doped graphene[12] versus n-doped graphene from alkali metal (Li) intercalation.[13, 14] Gold intercalation beneath single layer graphene on SiC(0001) was reported to p-dope the graphene layer.[15, 16] Theoretical studies demonstrated that modifications in the chemical composition of the buffer can lead to significant changes of the graphene bands, allowing for a fine tuning of the electronic structure of the system with band offsets up to 1.5 eV and even induce magnetism of the graphene.[17]

Alkali metal adsorption on graphene has been used extensively to study the effects of electron doping. Potassium deposition on graphene on SiC(0001) has enabled detailed studies of quasiparticles[18] and the discovery of plasmarons[19] in graphene. The additional dipole field from adsorbed potassium can also act to break the sublattice symmetry in graphene to induce bandgap opening.[20]

It is of interest to understand electron doping effects in alkali- intercalated graphene particularly because it could enable studies of correlated electron phenomena such as superconductivity or ferromagnetism in graphene due to the presence of extended van Hove singularities.[21] The intercalation of Ca was reported by McChesney et al. to result in electron doping of epitaxial graphene on SiC(0001) that moved the Dirac point an additional ~0.5 eV below the Fermi level.[21]

In this Letter we describe scanning tunneling microscope (STM) studies of the intercalation of sodium in epitaxial graphene on SiC(0001) at room temperature and the effects of subsequent annealing on the intercalation structures. In addition, we report electron doping effects inferred from the scanning tunneling spectroscopy (STS) study of image potential-derived surface states (IPS's). These states arise outside of a surface due to polarization of surface charge[22] and have recently been connected with the formation of interlayer states in graphite[23, 24] and used to understand screening and doping effects in epitaxial graphene in SiC(0001).[25] Changes in work function due to doping are often visible as corresponding shifts in the energy of IPS's[26] and they can be used to demonstrate doping effects in sodium intercalated epitaxial graphene. These results have been interpreted using electronic structure calculations from first principles that unambiguously distinguish the different locations of Na in the intercalation structures.

The SiC samples used in this study were chemical-mechanical polished (NovaSiC) 4H-SiC(0001) wafers. They were cleaned ex situ with acetone, methanol, and HF before introduction to an ultrahigh vacuum (UHV) system (Omicron). Using direct current heating the samples were annealed in UHV (base pressure $\sim 2\times 10^{-11}$ torr) to 1300 °C for 3 minutes to grow graphene. By choosing a lower annealing temperature (1000 °C for 10 minutes) of the SiC substrate we can prepare a surface with mainly bare buffer layer coexisting with small domains of graphene.

The samples were allowed to cool for several hours to approximately room temperature. Sodium was then deposited from an outgassed getter source (SAES). STM images were measured using a commercial instrument (Omicron) in constant current mode at room temperature with electrochemically-etched tungsten tips. Annealing experiments after sodium deposition were carried out using a calibrated tungsten heater in the sample manipulator and are expected to be accurate to about 20 °C. Scanning tunneling spectroscopy was performed in distance-versus-voltage ("z(V)") mode with constant current feedback engaged. The junction voltage is swept in a positive large bias interval where high local density of states is observed as an increased rate of tip retraction.[27]

Theoretical calculations were carried out using plane-wave, pseudopotential, density functional theory as implemented in the quantum-ESPRESSO software package[28] with the generalized gradient approximation (GGA) and the Perdew-Becke-Ernzerhof functional and a correction for dispersion forces (DFT(D)). Other computational details are the same as in Ref 17. All calculations include 1 Na atom per 8 graphene carbon atoms in a computational unit cell.[17]

At low coverages, sodium exposure on single layer graphene (SLG) results in the appearance of linear chains on the surface following the 3-fold symmetric directions of the SiC substrate as shown in Figure 1a. In Figure 1b, at higher submonolayer coverage, the chains become more densely packed and coalesce into islands which have the same 6×6 corrugation as the bare single layer graphene (Figure 1c). The STM images in Figure 1d clearly show these domains to be covered with a honeycomb graphene lattice. We infer that sodium is intercalated in between the buffer layer and the graphene layer. We abbreviate the compact domain in Figure 4c as SiC/B/Na/G where B stands for buffer layer and G for graphene layer. Judging from the apparent height and continuity of this intercalation structure, the islands are locally the same (e.g. SiC/B/Na/G) as the chains. A recent DFT study supports the idea that Na intercalation between the buffer layer graphene is preferred over Na adsorption on top of the graphene layer.[29] Sodium atoms completely avoid domains of BLG (marked in fig. 1b). Our experimental observation of intercalation in epitaxial single layer graphene is unexpected since Na does *not* form intercalation compounds on graphite without help from catalytic impurites.[30]

An additional intercalation structure observed after exposing epitaxial graphene to Na is illustrated in Figure 2a-2c. It exists in small domains immediately after deposition of Na at room temperature, but dominates the surface by heating of the sample after Na deposition. Figure 2a shows an area of a sample that has been annealed to approximately 180 °C that is now strikingly different from the SiC/B/Na/G layers in Figure 1. It exhibits a flat morphology with no sign of periodic 6×6 corrugation from the substrate. Instead, domains have randomly-distributed depressions and line defects. Figure 2b directly demonstrates the comparison between this surface and bare SLG and BLG.

The vanishing corrugation from the substrate is reminiscent of recent STM observations of hydrogen-exposed graphene on SiC where the hydrogen has intercalated at the SiC-buffer layer, decoupling the buffer layer to create a second graphene layer.[10] Figure 2c shows a close up of an intercalated domain after annealing of the sample where it is apparent that the surface is covered graphene but now with a triangular symmetry of the graphene honeycomb. This is strong evidence that we now have two layers of graphene sheets that are stacked in regular Bernal stacking (AB). We therefore conclude that the sodium has penetrated beneath the buffer layer and converted it into a second graphene layer just like the case for lithium,[13, 14] hydrogen,[8-10] oxygen,[11] and fluorine[12] intercalation. We abbreviate this intercalation structure SiC/Na/G/G.

To directly verify that the buffer layer can convert to a second graphene layer due to Na intercalation, we repeated the measurement with a surface prepared to consist of bare buffer layer. Images in Figures 3d-3f show such a surface after Na deposition and annealing to ~180 °C. The sodium has intercalated at the buffer layer-SiC interface across the whole surface. We label this geometry as SiC/Na/G. Its surface morphology is similar to the previously described structure SiC/Na/G/G where the 6×6 surface corrugation has vanished. A unique characteristic of this structure is the shape of the depressions. They have a distinct triangular shape (see figure 3d-3e) that is most likely related to the reduced buffer layer-substrate distance at high symmetry points that locally inhibits Na intercalation.[31] The STM image in Figure 3e proves that the buffer layer is decoupled since the graphene honeycomb lattice is clearly visible.

To further characterize the three intercalated structures, we measured their IPS's using STS and compared these to SLG. Our expectation is that the variation in energy of these states between different intercalation structures will reflect variations in the Na doping-induced changes in work function.[26] Figure 3 displays differentiated distance-voltage (dz/dV) spectra for intercalated structures relative to SLG (red curve, with single $n=1$ IPS peak defined as 0 eV) where the peak corresponds to the first member of the (odd[24]) Rydberg series that we refer to as *n=1*.

Since the *n=1* peak position of epitaxial graphene is sensitive to different tip states due to poor screening in the graphene sheet,[25] it is important to keep track of the tip state. We collected the STS data for intercalation structures SiC/Na/G, SiC/Na/G/G and SLG simultaneously (i.e. with the same tip state) as an internal reference. Intercalation structure SiC/B/Na/G was measured simultaneously with BLG and referenced to SLG using the known 0.14±0.02 eV difference in IPS energy between these two.[25]

For SiC/B/Na/G, the n=1 state is shifted down (purple curve) compared to bare SLG on SiC(0001). This shift can be attributed to a lowering of the work function as sodium electron-dopes the graphene through charge transfer. For SiC/Na/G/G, the n=1 IPS is shifted up again and is slightly lower in energy compared to SLG (green curve). This reversal in direction of work function can be explained by the hypothesis that Na atoms penetrate beneath the buffer layer so that their doping of both layers appears reduced compared to when sodium sits between buffer and graphene layer. This is analogous to the differences in substrate-induced doping between single layer and bilayer epitaxial graphene on SiC(0001).[32] For SiC/Na/G (blue curve) the downshift in n=1 IPS implies that electron doping is increased and work function decreased in a similar manner to SiC/B/Na/G. This similarity is expected since both these structures have Na intercalated directly under one graphene sheet.

Our interpretation of the location of intercalated Na atoms and associated electron doping effects of the three different intercalation structures can be substantiated using first principles electronic structure calculations. Figure 4 displays theoretical band structures for SLG (Figure 4a) along with SiC/B/Na/G (Figure 4b), SiC/Na/G/G (4c) and SiC/Na/G (4d). For SLG, Si dangling bonds at the SiC- buffer layer interface are responsible for a charge transfer which induces a shift of the Dirac Point to 0.49 eV below the Fermi level (red dashed line in Figure 4). For SiC/B/Na/G the Dirac point is shifted down further to 1.09 eV below Fermi level. This is consistent with an earlier calculation made for a similar geometry.[29] In addition, the shift is similar in size to the shift of the Dirac point reported for Ca intercalation by McChesney et al.[21]

Calculations for SiC/Na/G/G shown in Figure 4c demonstrate that the Na atoms between the buffer layer and SiC surface decouple the buffer layer. The electron doping for the remaining AB stacked graphene layers comprising SiC/Na/G/G is reduced compared to SiC/B/Na/G and a 0.29 eV gap opens around the Dirac point with mid gap located 0.65 eV below the Fermi level. For SiC/Na/G (Figure 4d), doping from intercalated Na results in a downshift of its Dirac point to 0.99 eV below the Fermi level. Importantly, the linear band structure around the Dirac point is maintained in this isolated, activated buffer, indicating that it is a true graphene sheet in accordance with the direct STM imaging in Figure 3f.

To compare IPS energy shifts with theory we used the results of *ab initio* calculations to extract work functions for each structure. Image 4e displays IPS n=1 position and calculated work function shift relative to the reference point of SLG. The trend of the calculated work function changes agrees with the STS measurements of the change in n=1 IPS peak position. The most important correspondence from the comparison in Figure 4e is the reversal in direction of work function change in going from SiC/B/Na/G to SiC/Na/G/G intercalation. This verifies the hypothesis that heating of the sample initiates Na atom intercalation *beneath* the buffer layer and not in direct contact with the upper graphene layer.

In summary, we observed the intercalation of sodium deposited on single layer graphene grown on SiC(0001). Contrary to graphite, sodium intercalates readily at the surface at room temperature and forms two different structures. At first, the sodium goes in between the top single graphene layer and the carbon buffer layer. Over time, or with annealing, sodium penetrates through the buffer layer converting this into a second graphene layer. Buffer layer decoupling by intercalation is directly observed by deposition of Na onto bare buffer layer followed by annealing.

First principles DFT calculations directly show the electronic structure and work function trends for these structures. The later compare favorably with measured shifts in IPS energies. These observations point out the rich possibilities for tailoring electronic structure by intercalation of graphene. The very strong electron doping reported here for Na intercalation could be valuable in the further exploration of superconductivity and magnetism in graphene.[21]  Moreover, the efficiency of the intercalation processes identified for Na suggests that intercalation of epitaxial graphene may be an even more versatile strategy for functional modification than it is for graphite.

Figure Captions:

Figure 1. (color online) STM images of Na intercalated between buffer and graphene layer, SiC/B/Na/G. (a) Low coverage Na chains (V=-2.18 V, 62 pA); (b) Intermediate coverage with chains and islands. (V=-2.0 V, I=50 pA); c) Na island in figure b (V=-2.3 V, I=50 pA); Graphene lattice of island in figure c (V=-0.55 V, I=350 pA).

Figure 2. (color online) STM images of Na intercalated at SiC interface after heating of sample: SiC/Na/G/G (a)-(c) and SiC/Na/G (d)-(f); (a) ~180 °C with full coverage SiC/Na/G/G (V=-1.93 V, I=200 pA); (b) ~530 °C showing an area of coexisting SLG, BLG, and SiC/Na/G/G (V=-1.7 V, I= 50 pA); (c) Graphene lattice on SiC/Na/G/G (V=-0.37 V, I=150 pA); (d) ~180 °C with full coverage SiC/Na/G (V=-2.6 V, I= 50 pA) , (e) Triangular depressions covering SiC/Na/G surface (V=-1.8 V, I= 50 pA); (f) Graphene lattice on SiC/Na/G (V=-0.14 V, I=0.11 nA).

Figure 3. (color online). Differentiated (dz/dV) distance-voltage spectra z(v) of the n=1 IPS for different intercalation structures relative to the n=1 IPS on SLG (red). All spectra are recorded with a constant current set-point of 50 pA.

Figure 4. (color online) Calculated electronic bands- and atomic structures for (a) SLG on top of SiC(0001); (b) SiC/B/Na/G (c) SiC/Na/G/G (Layers are AB-stacked); and (d) SiC/Na/G. Yellow atoms are C, black are Si, and red are Na. (e) Shift in measured n=1 IPS energy (blue squares) and shift in calculated work function from SLG (green circles) for each intercalation structure.

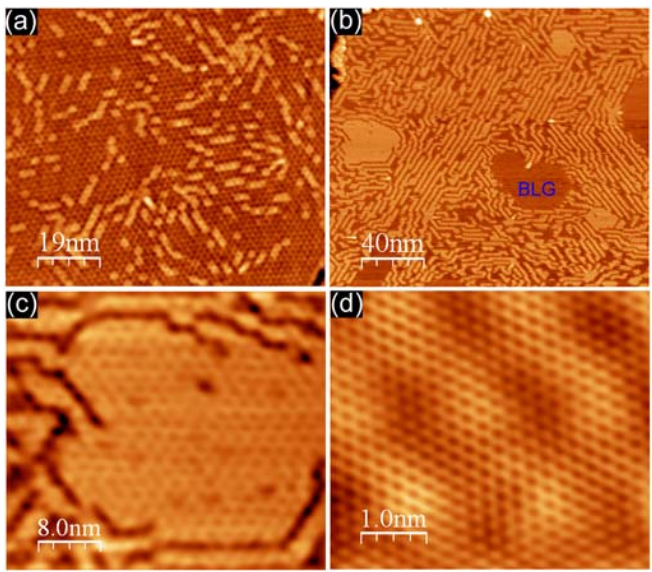

Figure 1, Sandin et al.

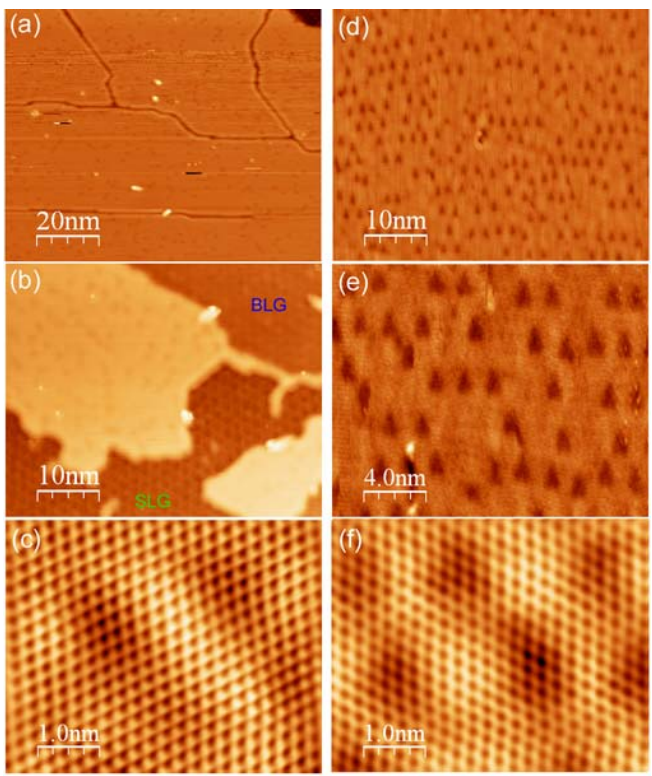

Figure 2, Sandin et al.

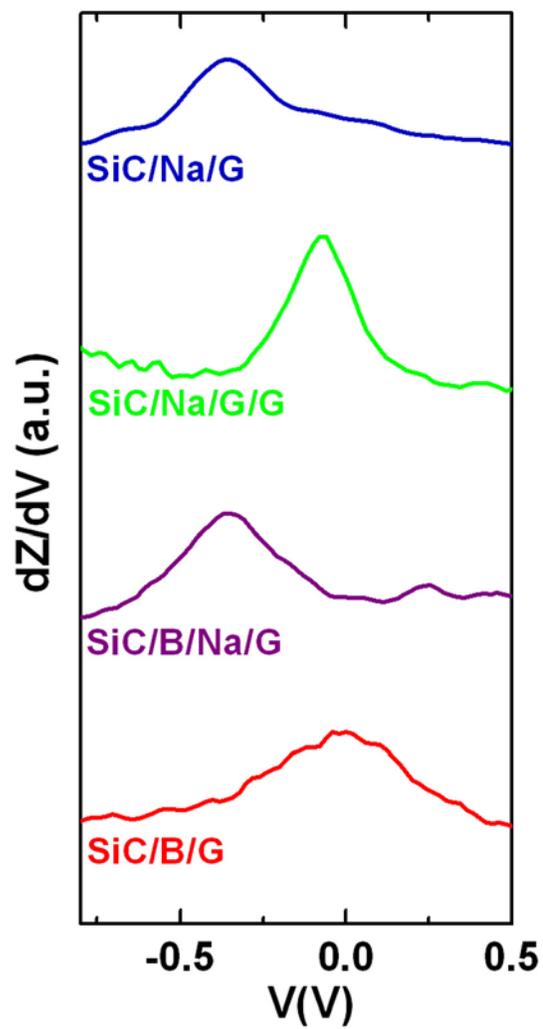

Figure 3, Sandin et al.

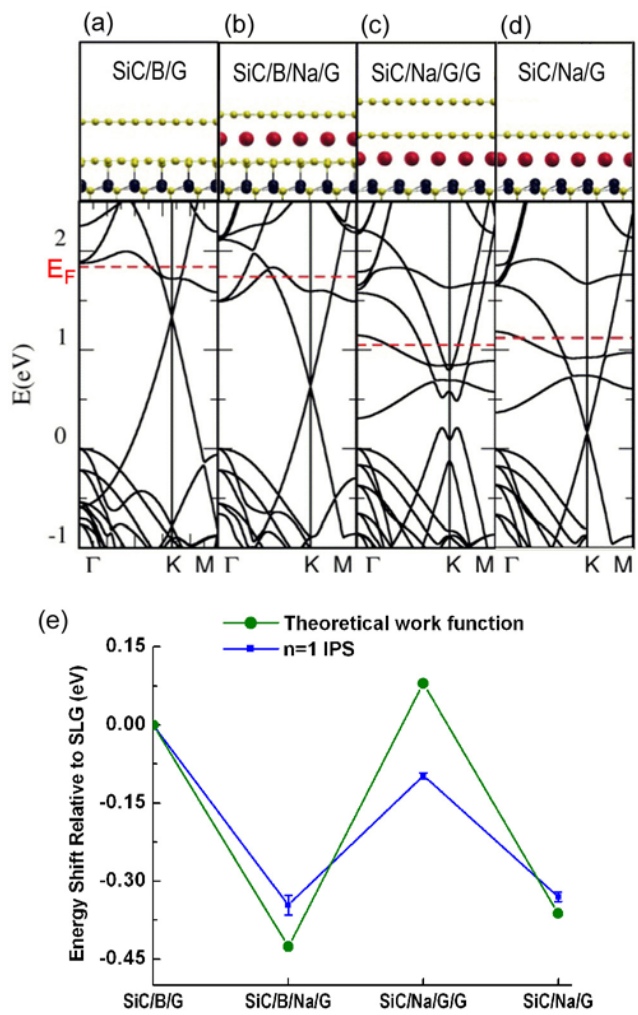

Figure 4, Sandin et al.